\documentclass[showpacs,aps,amssymb,floatfix,prd,amsmath,preprintnumbers]{revtex4}
\usepackage{epstopdf}
\usepackage{capt-of}
\usepackage{hyperref}

\usepackage{graphicx}  
\usepackage{dcolumn}   
\usepackage{bm}
\begin{document}
\input epsf.tex

\title{\bf Mixed fluid cosmological model in $f(R,T)$ gravity}

\author{
 Parbati
Sahoo\footnote{Department of Mathematics, National Institute of
Technology, Calicut, Kerala-673601 India,
India,  Email:  sahooparbati1990@gmail.com.}, Barkha Taori
\footnote{Department of Mathematics, Birla Institute of
Technology and Science-Pilani, Hyderabad Campus, Hyderabad-500078, India,
India,  Email: barkha.m.taori@gmail.com}, K. L.
Mahanta \footnote{ Department of Mathematics, C.V. Raman College
of Engineering, Bidya Nagar, Mahura, Janla, Bhubaneswar-752054,
India,  Email:  mahantakamal@gmail.com}. }

\affiliation{ }

\begin{abstract}
We construct Locally Rotationally Symmetric (LRS) Bianchi type-I
cosmological model in $f(R,T)$ theory of gravity when the source
of gravitation is the mixture of barotropic fluid and dark energy
(DE) by employing a time varying deceleration parameter (DP). We
observe through the behavior of the state finder parameters $(r,
s)$ that our model begins from the Einstein static era and goes to
$\Lambda$CDM era. The EoS parameter($\omega_d$) for DE varies from
phantom ($\omega<-1$) phase to quintessence ($\omega >-1$) phase
which is consistent with the observational results. It is found
that the discussed model can reproduce the current accelerating
phase of expansion of the universe.
\end{abstract}



\maketitle

\section{Introduction}

The latest cosmological observation detects the expansion of the universe as an accelerating rate \cite{Riess/1998, Perlmutter/1999}. This has led us to consider the exotic matter, dark energy, as a clarification and proof of this late time acceleration. DE existence has been supported by various observational data which includes – Cosmic Microwave Background (CMB) anisotropy \cite{Bernardis/2000,Bennett/2003,Hanany/2000}, Large Scale Structure (LSS) \cite{Spergel/2003, Tegmark/2004, MTegmark/2004}, Sloan Digital Sky Survey (SDSS)\cite{ Tegmark/2004, MTegmark/2004,Seljak/2005, Adelman/2006}, Wilkinson Microwave Anisotropy Probe \cite{Knop/2003} (WMAP) and Chandra X-ray  observatory \cite{Allen/2004}. A further investigation has established our universe composition as $73\%$ dark energy, $23\%$ dark matter and only $4\%$ as baryonic matter. DE is a scalar field of negative pressure with positive energy which serves as a means for reverse gravitational action \cite{Riess/2004,Tonry/2003}. This explains the shift from early time inflation to late time acceleration. \\

DE can be explained in two ways. The first one is by
choosing any of the exotic matter options viz. – quintessence
\cite{Martin/2008,Wetterich/1988,Ratra/1988}, phantom
\cite{Nojiri/2006}, $k$- essence
\cite{Chiba/2000,Armendariz/1999,Armendariz/2000}, tachyons
\cite{Padmanabhan/2002}, quintom
\cite{Nojiri/2004}, chaplygin gas
\cite{Srivastava/2005,Bertolami/2004,Bento/2002,Bilic/2002,Avelino/2003,Kamenshchik/2001},
chameleon \cite{Khoury/2004}, cosmological nuclear energy
\cite{Gupta/2010} and cosmological constant
\cite{Overduin/1998,Sahni/2000,Komatsu/2009,Kachru/2003}. Out of
these, the approach of cosmological constant is the simplest and
most general to explain the acceleration but it incorporates the
problems related to cosmic coincidence and fine-tuning
\cite{Sahni/2000,Weinberg/1989,Peebles/2003}. However, these
choices are insufficient to explain the mystery of dark energy
completely. DE model is outlined using the Equation of
State (EoS) parameter $\omega$ which defined as in terms of pressure
and energy density such that; $\omega(t) = \frac{p}{\rho}$. This parameter
need not be a constant \cite{Carroll/2003}. It can be parametrized as in terms
of time or scale factor ($a$) or redshift ($z$). \\

The second way to explain DE is by modifying the theory of
gravity. These theories serve natural gravitational alternatives
to DE and attempt to justify current acceleration.  Various
modified theories are $f(R), f(G), f(T)$ and $f(R,T)$. The
generalization of the Lagrangian in Einstein-Hilbert action where,
function $f(R)$ is used instead of $R$, Ricci scalar gives $f(R)$
theory of gravity \cite{SNojiri/2011}. This theory serves as a
consolidation of early time inflation with late time acceleration.
The model handles higher order curvature invariants as a function
of $R$. A further generalization of $f(R)$ gravity theory yields
$f(R, T)$ theory which was originally introduced by Harko et al.
\cite{Harko/2011}. The authors considered Lagrangian density as a
function $f(R, T)$, where $T$ denotes the trace of energy momentum
tensor. This model, contrast to other theories, discuss matter and
geometry coupling. This results in source term independence, where
source term is the matter stress energy tensor variant. They claim
that cosmic acceleration is also a result of matter content
besides geometrical input. Thereafter, many researchers are
interested to do more investigations of this theory through
various aspects \cite{Houndjo/2012, Jamil/012, Alvarenga/2013,
Houndjo/2014, Sahoo/017,PKSahoo/2017, Sharif/12, Alvarenga/013}.
In \cite{Shabani/2013} FLRW cosmological model has been studied in
the framework of $f(R,T)$ gravity through phase space analysis. We
can see in refs.\cite{Sahoo/2016, Sahoo/2017,Sahoo/0017} it has
been studied for different matter components. Recently, V. Fayaz
et al. \cite{Fayaz/2017}  studied Bianchi-I space-time in this
theory where they regenerated $f(R,T)$ function using holographic
dark energy. They reproved that the rate of evolution of the
anisotropic universe is greater than that of FRW and $\Lambda$CDM
model.

Yadav et al. \cite{Yadav/2011} discussed the DE model in Bianchi
type-III universe with constant DP ($q$). The
EoS parameter $\omega$ is established as a time dependent factor
in the respective case. Naidu et al. examined spatially homogeneous
and anisotropic Bianchi type II \cite{Reddy/2012} and III
\cite{RLNaidu/2012} models based on Saez and Ballester theory.
Based on the same theory, the authors also investigated Bianchi
type V \cite{Naidu/2012} model with variable $\omega$ and constant
$q$. Kotambkar et al. \cite{GPSingh/2014} constructed anisotropic
Bianchi type I model with bulk viscosity and quintessence and
discussed various physical properties of the model. Singh et al.
\cite{Singh/2014} examined Bianchi type II model for a perfect
fluid source in $f(R,T)$ gravity. The solutions were obtained
using the power law relation between mean Hubble parameter
($H(t)$) and average scale factor ($a(t)$). The same conditions
were worked upon by Reddy et al. \cite{Reddy/2013} using the
special law of variation for Hubble's parameter given by Berman
\cite{Berman/1983}. The special law generates constant
DP which implies exclusion of open universes.
Samanta \cite{sam13} constructed a model of the universe filled
with dark energy from a wet dark fluid in $f(R,T)$ gravity.
Samanta and Dhal \cite{samanta13} studied Bianchi type- V universe
with a binary mixture of perfect fluid and dark energy in $f(R,T)$
gravity. Sahoo and Mishra \cite{sah14} investigated Kaluza-Klein
dark energy model in the form of wet dark fluid in $f(R,T)$
gravity. Singh and Sharma \cite{Singh/2014} constructed Bianchi type-II
dark energy cosmological model with variable EoS parameter in
$f(R,T)$ gravity. By considering constant DP
they obtained two models of the universe, namely, power law model
and exponential model. Yadav et al \cite{yad15} obtained dark
energy dominated universe in $f(R,T)$ gravity with hybrid law
expansion. Rao et al. \cite{Rao/2015} applied the same law to
investigate 5-dimensional Kaluza-Klein space time in the existence
of anisotropic dark energy in $f(R, T)$ gravity. Bishi
\cite{bis16} studied Bianchi type-III dark energy model in
$f(R,T)$ gravity with variable DP. Chaubey et
al. \cite{cha16} considered general class of Bianchi cosmological
models in $f(R,T)$ gravity with the dark energy in the form of
standard and modified Chaplygin gas. Sahoo \cite{sah17} considered
Kaluza-Klein universe filled with wet dark fluid in $f(R,T)$
gravity and obtained the exact solutions from a time varying DP.   \\

In this work, we use both the approach concurrently. That is we
considered the source of gravitational matter as a mixture of
perfect fluid and dark fluid in a modified theory called $f(R,T)$
theory. This type of simultaneous use of both the approach have
already been used by the authors \cite{baf17} and \cite{sha17}.\\

The work being organized in the following manner: In Section-I,
Introduction and motivations from the literature are briefly
elaborated. Section-II contains the basic formalism of $f(R,T)$
gravity general field equations. The solution of the field
equation for LRS Bianchi type-I metric by employing time varying
DP are presented in Section-III. At last the
Physical behavior of the model and conclusions are outlined in
Section-IV and Section-V respectively.

\section{The $f(R,T)=R+2f(T)$ gravity}
 Field equation for $f(R,T)$ gravity can be formulated from the Hilbert-Einstein in the following manner.
\begin{equation} \label{eqn1}
S=\frac{1}{16\pi G}\int \sqrt{-g}f(R,T)d^{4}x+\int \sqrt{-g} L_{m}d^{4}x,
\end{equation}%
where $L_{m}$ is the matter Lagrangian density, $g$ is the determinant of the metric tensor $g_{ij}$, $R$ is Ricci scalar  and $T$ is the trace of energy-momentum tensor $T_{ij}$. The energy-momentum tensor $T_{ij}$ is defined as
\begin{equation}\label{eqn2}
T_{ij}=-\frac{2}{\sqrt{-g}}\frac{\delta (\sqrt{-g}L_{m})}{\delta g^{ij}}.
\end{equation}%
Here, Instead of considering the derivative of matter Lagrangian, we have assumed that the matter Lagrangian $L_{m}$ depends only on the metric components. Such as
\begin{equation}\label{eqn3}
T_{ij}=g_{ij}L_{m}-\frac{\partial L_{m}}{\partial g^{ij}}.
\end{equation}%
The $f(R,T)$ gravity field equations are obtained from the eqn. (\ref{eqn1}) by varying the action $S$ with respect to metric component. It is given as
\begin{equation}\label{eqn4}
f_{R}(R,T)R_{ij}-\frac{1}{2}f(R,T)g_{ij}+(g_{ij}\Box -\nabla _{i}\nabla
_{j})f_{R}(R,T)=8\pi T_{ij}-f_{T}(R,T)T_{ij}-f_{T}(R,T)\Theta _{ij},
\end{equation}%
where
\begin{equation}\label{eqn5}
\Theta _{ij}=-2T_{ij}+g_{ij}L_{m}-2g^{lm}\frac{\partial ^{2}L_{m}}{\partial
g^{ij}\partial g^{lm}},
\end{equation}
and  $f_{R}(R,T)=\frac{\partial f(R,T)}{\partial R}$, $f_{T}(R,T)=\frac{%
\partial f(R,T)}{\partial T}$, $\Box \equiv \nabla ^{i}\nabla _{i}$, where $%
\nabla _{i}$ is the covariant derivative.\newline

 To construct different kind of cosmological models according to the choice of matter source, Harko et al. \cite{Harko/2011} constructed three types of $f(R,T)$ gravity as follows
\begin{equation}\label{eqn6}
f(R,T)=\left\{
\begin{array}{lcl}
R+2f(T) &  &  \\
f_{1}(R)+f_{2}(T) &  &  \\
f_{1}(R)+f_{2}(R)f_{3}(T) &  &
\end{array}%
\right.
\end{equation}%
The general field equation for first frame of $f(R,T)=R+2f(T)$ gravity is given as \\
\begin{equation}\label{eqn7}
R_{ij}-\frac{1}{2}Rg_{ij}=8\pi T_{ij}-2f'(T)T_{ij}-2f'(T)\Theta_{ij}+f(T)g_{ij}.
\end{equation}

\section{Field equations and Solutions}
We consider the spatially homogeneous LRS Bianchi type-I metric as
\begin{equation}\label{eqn8}
ds^{2}=dt^{2}-A^{2}(dx^{2}+dy^{2})-B^2 dz^{2},
\end{equation}
where $A$, $B$ are functions of cosmic time $t$ only.\\
The stress-energy momentum tensor is in the form
\begin{equation}\label{eqn9}
T_{j}^{i}=T_{(m)j}^{i}+T_{(de)j}^{i},
\end{equation}
where $T_{(m)j}^{i}$ and $T_{(de)j}^{i}$ are energy momentum tensor of perfect fluid and dark energy respectively. These are given by
\begin{equation}\label{eqn10}
T_{(m)j}^{i}=diag[\rho_{m},-p_{m},-p_{m},-p_{m}],
\end{equation}
 and
\begin{equation}\label{eqn11}
T_{(de)j}^{i}=diag[\rho_{d},-p_{d},-p_{d},-p_{d}],
\end{equation}

where $p_{m}$, $\rho_{(m)}$ are pressure and energy density for
perfect fluid and $p_{d}$, $\rho_{d}$ are pressure and the energy
density for dark energy components respectively.

The field eqn. (\ref{eqn7}) with $f(T)=\alpha T$, where $\alpha$ is an
arbitrary constant, becomes
\begin{equation}\label{eqn12}
R_{ij}-\frac{1}{2}Rg_{ij}=(8\pi+2\alpha) T_{ij}+(2\alpha p+\alpha T)g_{ij}.
\end{equation}
In the framework of $f(R,T)$ gravity, in the term $(2\alpha p+\alpha T)$, $p$ is the isotropic pressure and $T$ is the trace of energy-momentum tensor.
According to \cite{poplawski06} and \cite{poplawski006} the trace of energy momentum tensor is of isotropic pressure and energy density i.e. $T=\rho-3p$.\\
The field eqn. (\ref{eqn12}) for the line element (\ref{eqn8}) is given as
\begin{eqnarray}
-\dot{H_1}-H_1^2-\dot{H_2}-H_2^2-H_1H_2=(8\pi+2\alpha)(p_m+p_d)-\alpha (\rho_m-p_m),\label{eqn13}\\
-2\dot{H_1}-3H_1^{2}=(8\pi+2\alpha)(p_m+p_d)-\alpha (\rho_m-p_m),\label{eqn14}\\
2H_1H_2+H_1^2=(8\pi+2\alpha)(\rho_m+\rho_d)+\alpha (\rho_m-p_m).\label{eqn15}
\end{eqnarray}
Here $H_1=\frac{\dot{A}}{A}$, $H_2=\frac{\dot{B}}{B}$ and the over
dot represent derivatives with respect to cosmic time $t.$ We have
six unknowns $H_1$, $H_2$, $\rho_m$, $\rho_d$, $p_m$, \& $p_d$ and three
equations. In order to obtain the exact solution, we have
assumed in first step the Bianchi identity $ G^{ij}_{;j}=0$ as it is followed from the definition of the Einstein tensor $ G_{ij}$ and $R_{ij}$ \cite{Misner/1970}. From which we have obtained the following relation. \\
\begin{equation}\label{eqn16}
\dot{\rho_m}+3(1+\omega_m)\rho_m H=0,
\end{equation}
where $ H=\frac{\dot{a}}{a}=\frac{1}{3} (\frac{2\dot{A}}{A} +
\frac{\dot{B}}{B})$ is mean Hubble parameter,
$a=(A^2B)^{\frac{1}{3}}$ is the average scale factor and
$\omega_m=\frac{p_m}{\rho_m}$ is EoS parameter of perfect fluid
considered as a constant \cite{akarsu10}. From eqn. (\ref{eqn16}) we
obtain the value of $\rho_m$ as
\begin{equation}\label{eqn17}
\rho_m=c_1 a^{-3(1+\omega_m)},
\end{equation}
where $c_1$ is an integration constant. Following \cite{Chawla13}
we have considered the time varying DP of the
form
\begin{equation}\label{eqn18}
q=-\frac{a\ddot{a}}{\dot{a}^2}=b(t),
\end{equation}
This geometric parameter has vital role in the description of the evolution of the universe, which defines the phase transition of the universe from past decelerating expansion to the recent accelerating one. Thus, it is well motivated to consider a time-dependent DP $q$ is due to the fact that the universe exhibits phase transitions, as revealed by the cosmic observations of SNe Ia. Also, the transitional phase of the universe can be determined by the signature flipping nature of DP, i.e. positive DP defines decelerating phase and negative sign of DP represents the accelerating phase of late universe. Thus, the choice of time-dependent DP is physically reliable for cosmological models.
The expression of $q$ given in eqn. (\ref{eqn18}) can be written  as
 \begin{equation}\label{eqn19}
\frac{\ddot{a}}{a}+b\frac{\dot{a}^2}{a^2}=0.
\end{equation}
By assuming $b=b(a)$ or $b=b(a(t))$, the general solution of (\ref{eqn19}) is given as
\begin{equation}\label{eqn20}
\int e^{\int \frac{b}{a}}da=t+\textit{d},
\end{equation}
where $\textit{d}$ is a constant of integration. In order to derive the solution (\ref{eqn20}), without any loss of generality, one can choose $\int \frac{b}{a}da=\ln f(a)$, and eqn. (\ref{eqn20}) yields 
\begin{equation}\label{eqn21}
\int f(a)da=t+\textit{d}.
\end{equation} 
In this eqn. (\ref{eqn21}), the arbitrary function $f(a)$ can be chosen in such a way that, it will provide a physically viable and observationally consistent cosmological model. Thus, $f(a)$ is considered as 
\begin{equation}\label{eqn22}
f(a)=\frac{na^{n-1}}{\beta \sqrt(1+a^{2n})},
\end{equation} 
where $\beta$ is an arbitrary constant and $n$ is a positive constant. Using eqn. (\ref{eqn22}) in eqn. (\ref{eqn21}) and taking $\textit{d}=0$, we obtain the following exact solution is
\begin{equation}
a(t)=[\sinh(\beta t)]^{\frac{1}{n}}. 
\end{equation} 
Then the directional scale factors $A$ and $B$ are derived from the relation $V=a^3$ as follows  
\begin{eqnarray}
A(t)=[\sinh(\beta t)]^{\frac{1}{2n}},\label{eqn23}\\
B(t)=[\sinh(\beta t)]^{\frac{2}{n}}.\label{eqn24}
\end{eqnarray}
 Now the line element (\ref{eqn8}) can be rewritten as
\begin{equation}\label{eqn25}
ds^{2}=dt^{2}-[\sinh(\beta t)]^{\frac{1}{n}}(dx^{2}+dy^{2})-[\sinh(\beta t)]^{\frac{4}{n}} dz^{2}.
\end{equation}
From equations (\ref{eqn14}) and (\ref{eqn15}), the values of $p_d$, $\rho_d$ and $\omega_d$ are obtained as

\begin{equation}\label{27}
\rho_d=\frac{c_1 (\alpha  (\omega_m -3)-8 \pi ) \left(\left(\frac{1}{z+1}\right)^n\right)^{-\frac{3 (\omega_m +1)}{n}}+\frac{9 \beta ^2 \left(\left(\frac{1}{z+1}\right)^{-2 n}+1\right)}{4 n^2}}{2 (\alpha +4 \pi )},
\end{equation}

\begin{equation}\label{26}
p_d=\frac{c_1 (-3 \alpha  \omega_m +\alpha -8 \pi  \omega_m ) \left(\left(\frac{1}{z+1}\right)^n\right)^{-\frac{3 (\omega_m +1)}{n}}-\frac{\beta ^2 \left(\frac{1}{z+1}\right)^{-2 n} \left(3 \left(\frac{1}{z+1}\right)^{2 n}-4 n+3\right)}{4 n^2}}{2 (\alpha +4 \pi )},
\end{equation}

\begin{equation}\label{eqn28}
\omega_d=-\frac{4 c_1 n^2 (\alpha  (3 \omega_m -1)+8 \pi  \omega_m ) \left(\frac{1}{z+1}\right)^{2 n}+\beta ^2 \left(3 \left(\frac{1}{z+1}\right)^{2 n}-4 n+3\right) \left(\left(\frac{1}{z+1}\right)^n\right)^{\frac{3 (\omega_m +1)}{n}}}{9 \beta ^2 \left(\left(\frac{1}{z+1}\right)^{2 n}+1\right) \left(\left(\frac{1}{z+1}\right)^n\right)^{\frac{3 (\omega_m +1)}{n}}-4 c_1 n^2 (8 \pi -\alpha  (\omega_m -3)) \left(\frac{1}{z+1}\right)^{2 n}}.
\end{equation}

The variation of energy density, pressure and equation of state
(EoS) parameter with cosmic time t are shown in the following
figures. In Fig. \ref{fig1}, 
the energy density $\rho_d$ is a positive decreasing function of time and tends to zero at $t$ tends to $\infty$. Fig. \ref{fig-3}  represents the graphics for pressure $p$, which is a negative increasing function of time and tends to zero at $t$ tends to $\infty$. As per the observation, the negative pressure
is due to DE in the context of accelerated expansion of the universe. Hence, the behavior of pressure in our model agrees with this observation. In Fig. \ref{fig-5}, the EoS parameter lies in the accelerated phase dominated by DE era. From Fig. \ref{fig-5}, one can observe that the EoS parameter shows a transitional behavior. In Figs. \ref{fig-2}, \ref{fig-4}, and \ref{fig-6}, the variation of energy density, pressure and equation of state
(EoS) parameter with redshift parameter $z$ are depicted respectively, and it provides the reliability of the model.

\begin{figure}[h!]
\minipage{0.48\textwidth}
\includegraphics[width=78mm]{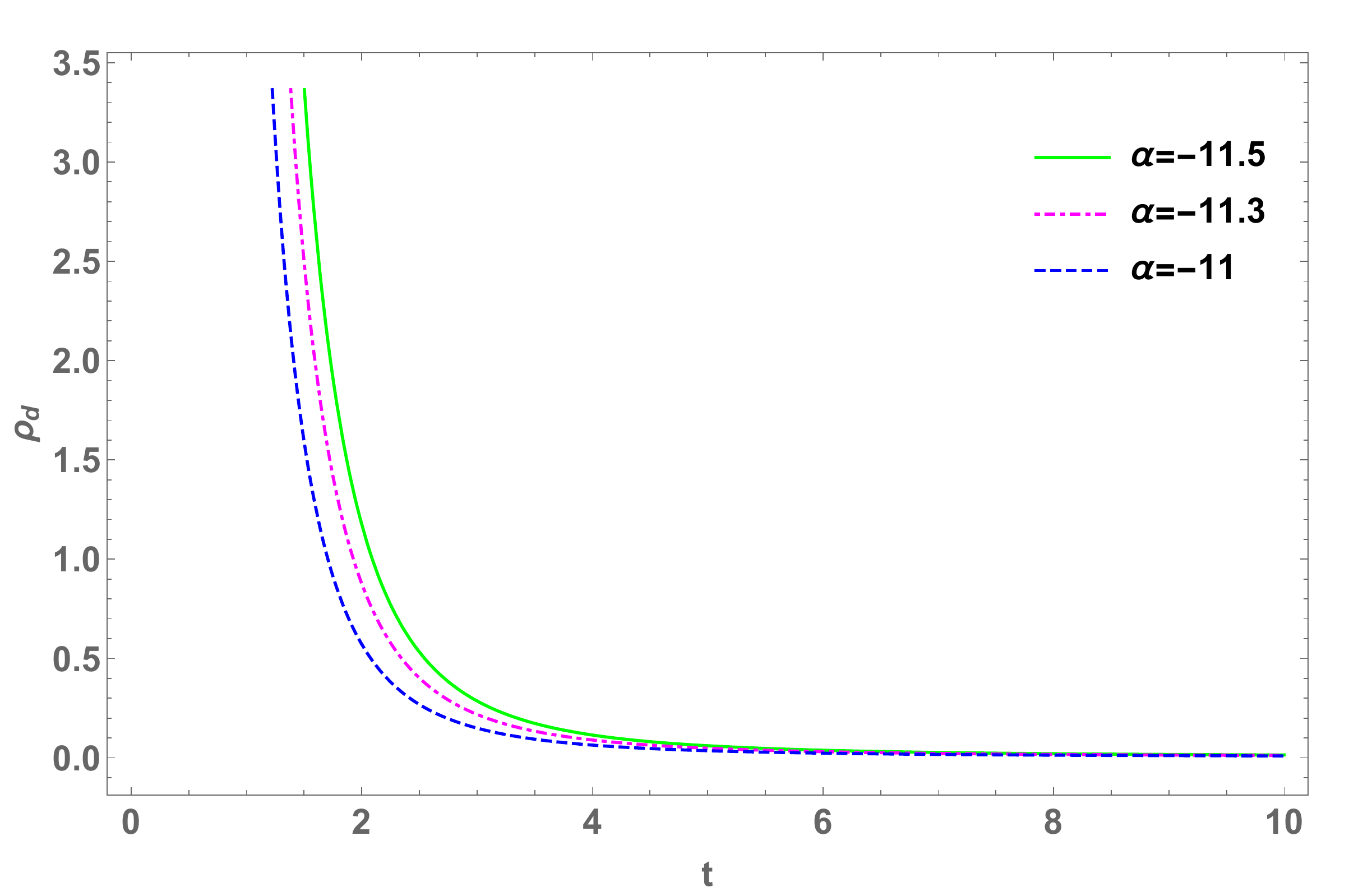}
  \caption{Variation of $\rho$ against time with $\beta=0.09804$, $n=1.15$, $c_1=0.001$, $\omega_m=0.5$.}\label{fig1}
\endminipage\hfill
\minipage{0.48\textwidth}
\includegraphics[width=78mm]{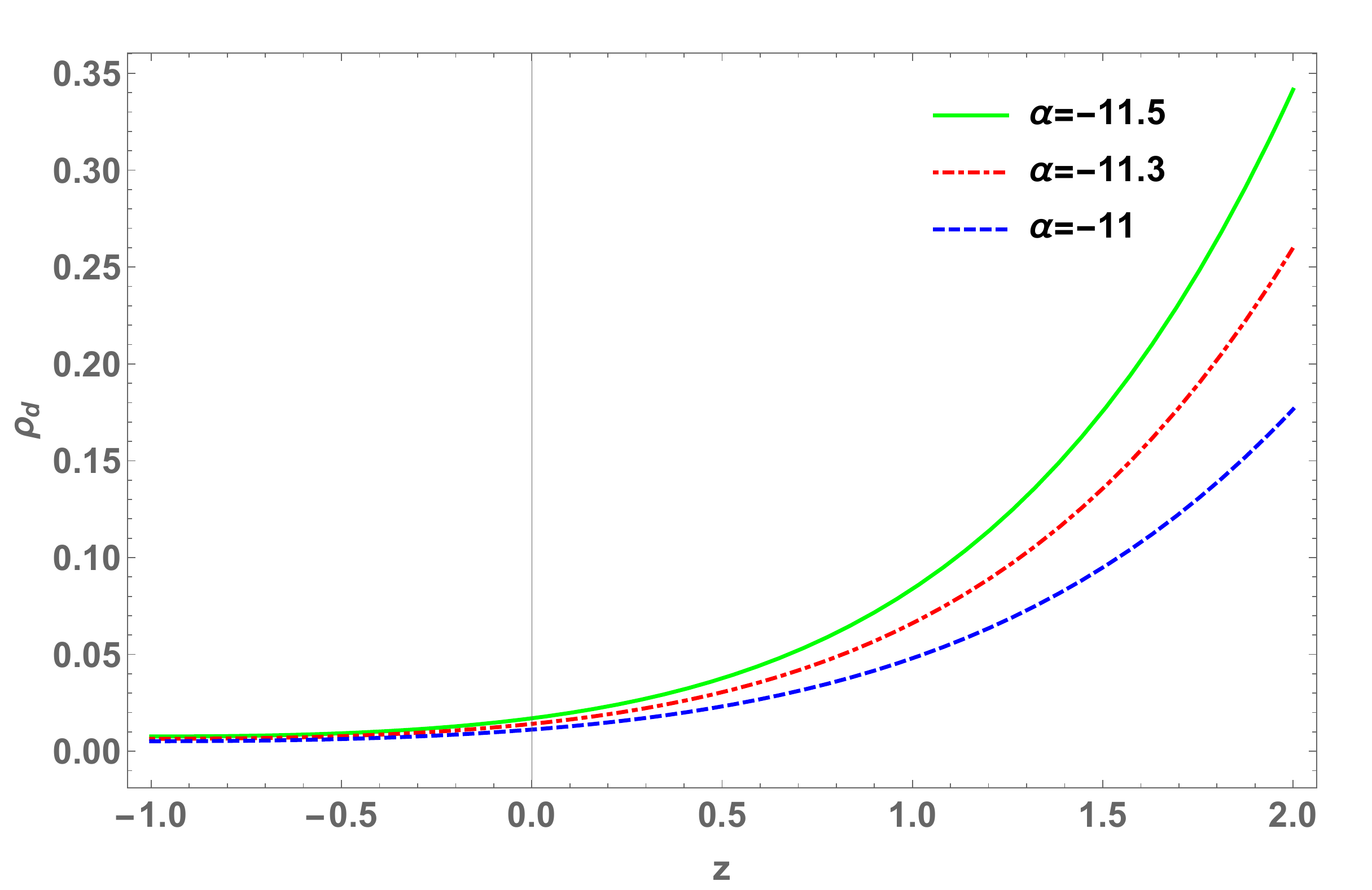}
  \caption{Variation of $\rho$ against $z$ with $\beta=0.09804$, $n=1.15$, $c_1=0.001$, $\omega_m=0.5$.}
\label{fig-2}
\endminipage
\end{figure}

\begin{figure}[h!]
\minipage{0.48\textwidth}
\includegraphics[width=78 mm]{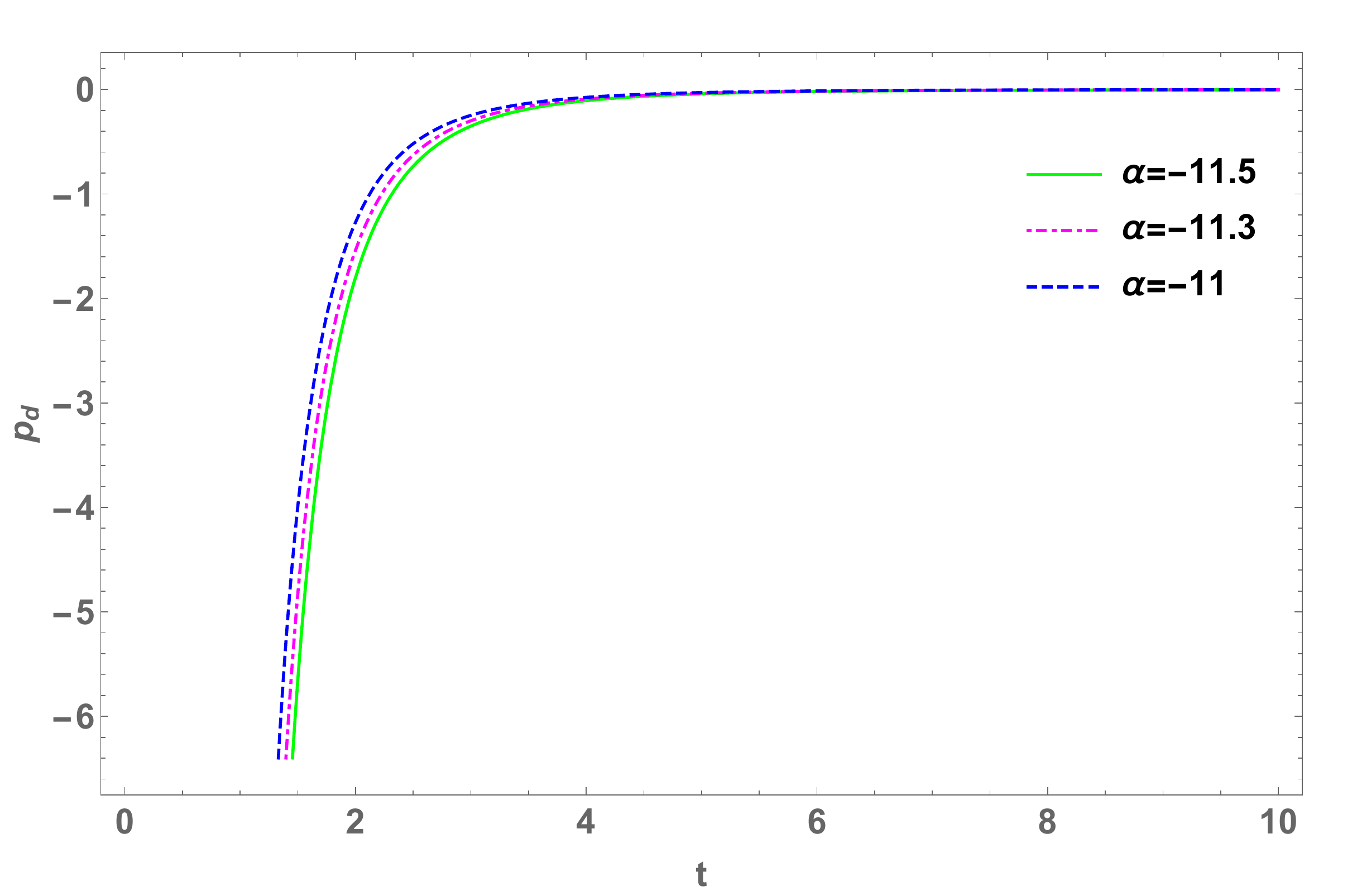}
  \caption{Variation of $p$ against time with $\beta=0.09804$, $n=1.15$, $c_1=0.001$, $\omega_m=0.5$.}\label{fig-3}
\endminipage\hfill
\minipage{0.48\textwidth}
\includegraphics[width=78 mm]{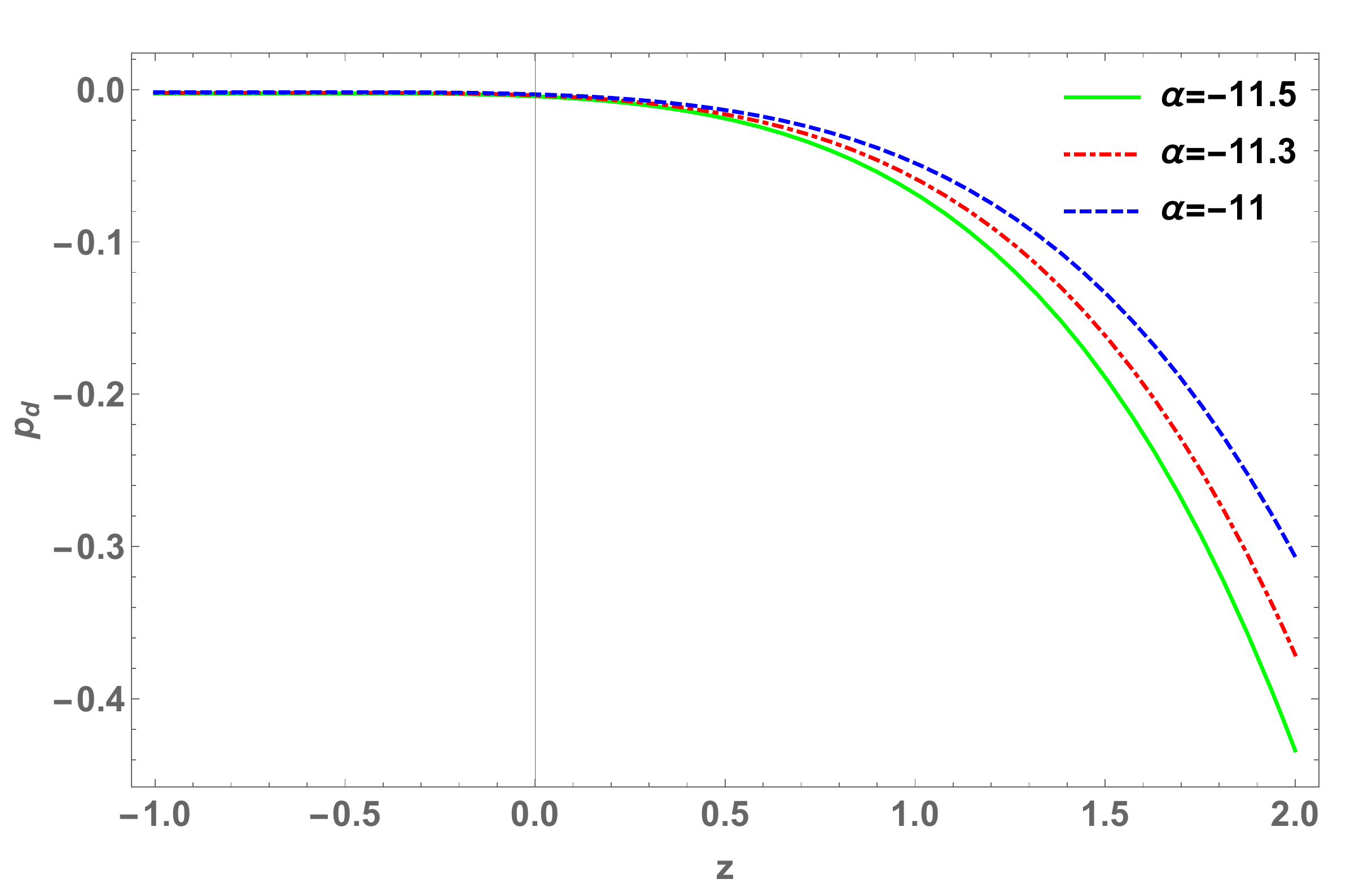}
  \caption{Variation of $p$ against $z$ with $\beta=0.09804$, $n=1.15$, $c_1=0.001$, $\omega_m=0.5$.}\label{fig-4}
\endminipage
\end{figure}

\begin{figure}[h!]
\minipage{0.48\textwidth}
\includegraphics[width=78 mm]{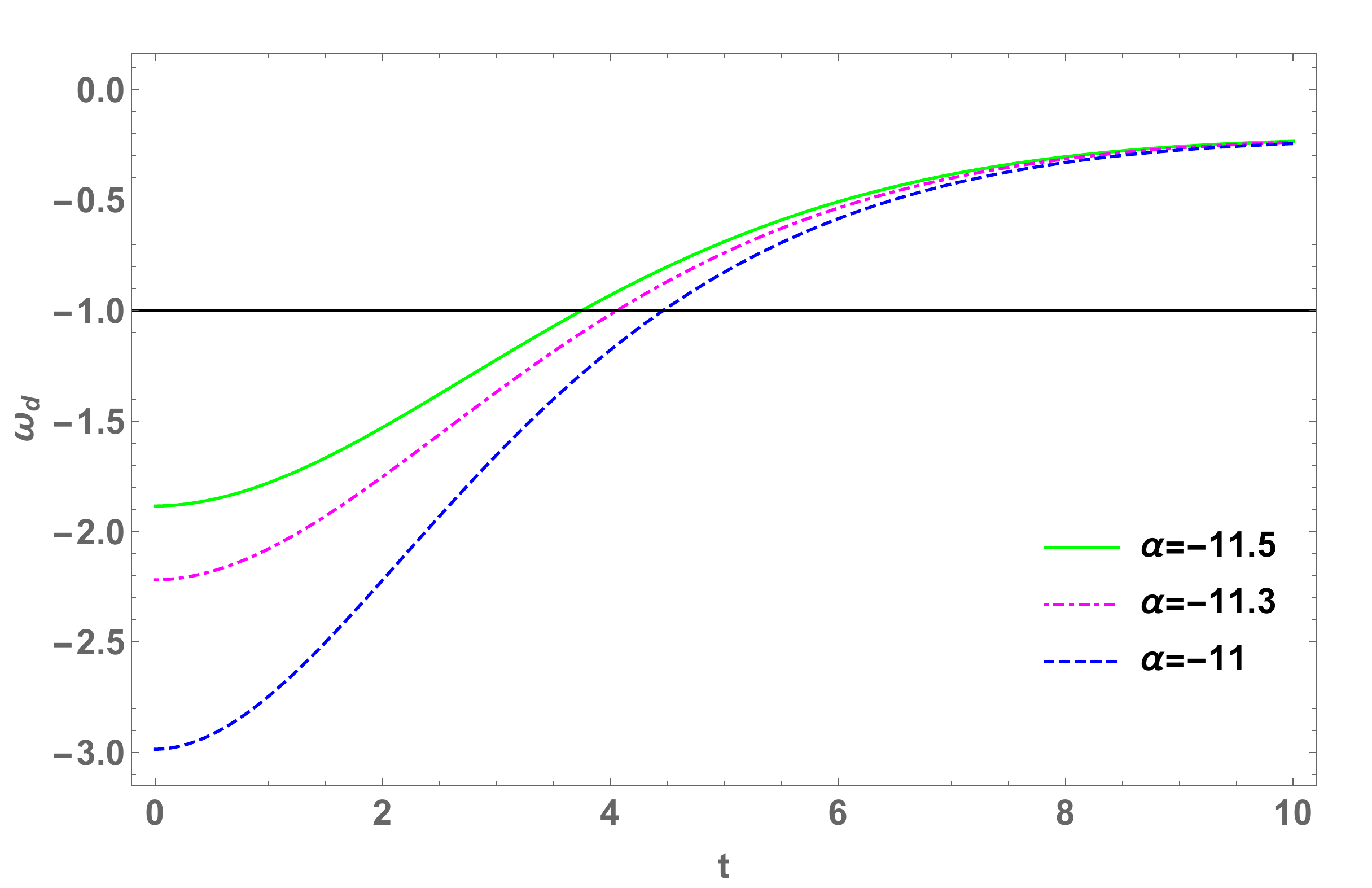}
  \caption{Variation of Eos Parameter against time with $\beta=0.09804$, $n=1.15$, $c_1=0.001$, $\omega_m=0.5$.}\label{fig-5}
\endminipage\hfill
\minipage{0.48\textwidth}
\includegraphics[width=78 mm]{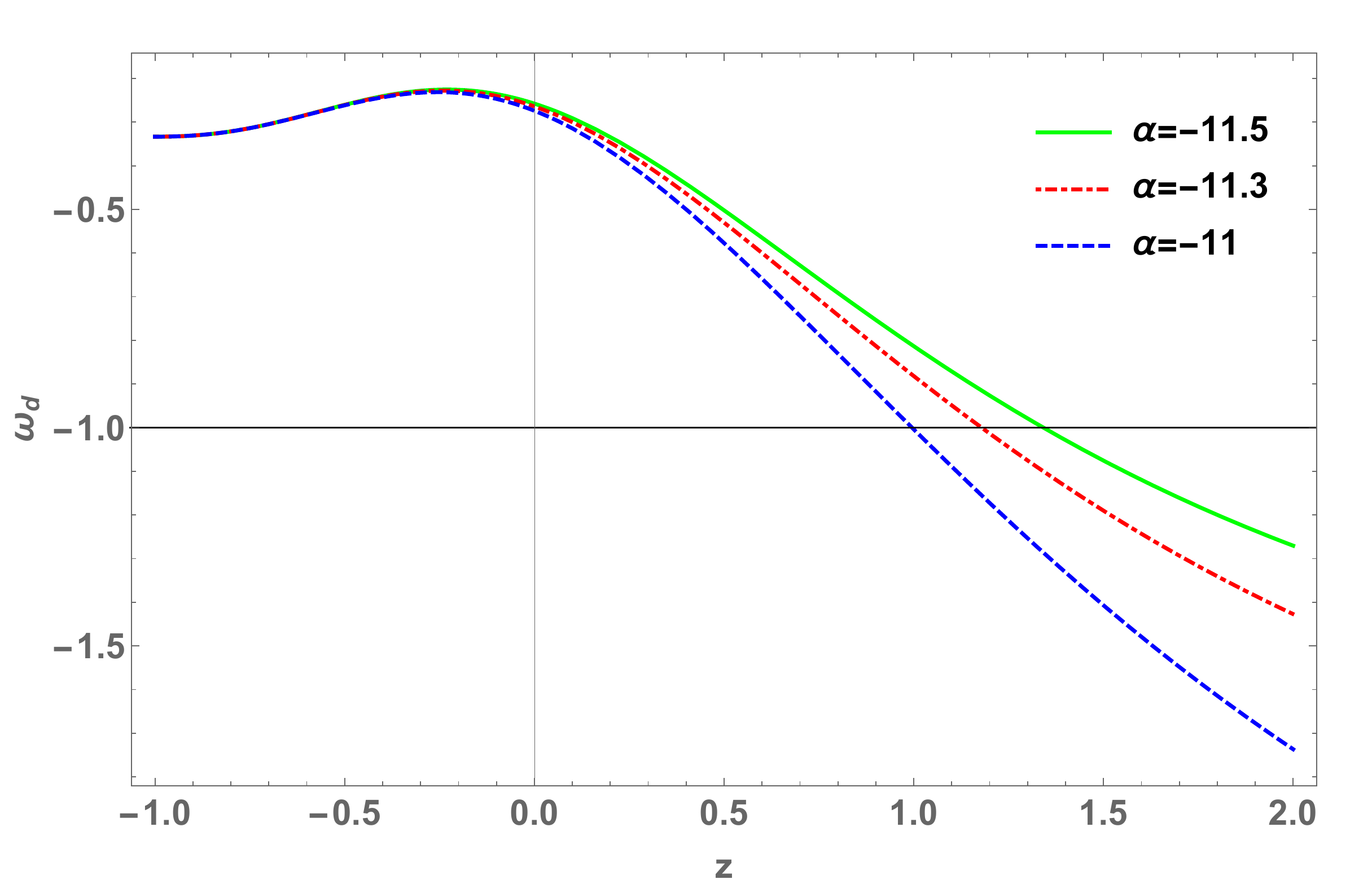}
  \caption{Variation of Eos Parameter against $z$ with $\beta=0.09804$, $n=1.15$, $c_1=0.001$, $\omega_m=0.5$.}\label{fig-6}
\endminipage
\end{figure}

\section{Physical properties of the models}
The values of various physical parameters viz. DP, energy density of perfect fluid ($\rho_m$), mean Hubble parameter ($H$), expansion scalar ($\theta$), shear scalar ($\sigma^2$) and mean anisotropic parameter ($A_m$) for the model are obtained as

 \begin{equation}\label{eqn29}
q=n \text{sech}^2(\beta  t)-1=\frac{n}{\left(\frac{1}{z+1}\right)^{2 n}+1}-1,
\end{equation}
\begin{equation}\label{eqn30}
\rho_m=c_1 \sqrt[n]{\sinh (\beta  t)}^{-3 \omega -3}=c_1 \left(\beta  \left(\frac{1}{z+1}\right)^n\right)^{\frac{-3 \omega_m -3}{n}},
\end{equation}
 \begin{equation}\label{eqn31}
 H=\frac{\beta  \coth (\beta  t)}{n}=\frac{\beta  \left(\frac{1}{z+1}\right)^{-n} \sqrt{\left(\frac{1}{z+1}\right)^{2 n}+1}}{n},
 \end{equation}
 \begin{equation}\label{eqn32}
 \theta=3H=\frac{3\beta  \coth (\beta  t)}{n}=\frac{3\beta  \left(\frac{1}{z+1}\right)^{-n} \sqrt{\left(\frac{1}{z+1}\right)^{2 n}+1}}{n},
 \end{equation}
 \begin{equation}\label{eqn33}
 \sigma^2=\frac{3 \beta ^2 \coth ^2(\beta  t)}{4 n^2}=\frac{3 \beta ^2 \left(\frac{1}{z+1}\right)^{-2 n} \left(\left(\frac{1}{z+1}\right)^{2 n}+1\right)}{4 n^2},
 \end{equation}
  \begin{equation}\label{eqn34}
 A_m=\frac{3}{2}.
 \end{equation}
 
 One of the important quantities for the dynamical description of the universe is known as state finder pair or $r-s$ parameter. It helps to study the coincidence between obtained model with $\Lambda$CDM model. For flat $\Lambda$CDM model, the value of state
   finder pair yields as $\{r,s\}=\{1,0\}$ \cite{Feng/2008}. The values of the $r-s$ parameter of our model becomes
 \begin{equation}\label{eqn35}
 r= \frac{\dot{\ddot{a}}}{aH^3}= n (2 n-3) \text{sech}^2(\beta  t)+1=\frac{n (2 n-3)}{\left(\frac{1}{z+1}\right)^{2 n}+1}+1,
 \end{equation}
 \begin{equation}\label{eqn36}
 s=\frac{r-1}{3(q-\frac{1}{2})}=\frac{n (2 n-3) \text{sech}^2(\beta  t)}{3 \left(n \text{sech}^2(\beta  t)-1.5\right)}=\frac{2 (3-2 n) n}{\left(\frac{1}{z+1}\right)^{2 n}-6 n+1}.
 \end{equation}

 \begin{figure}
\includegraphics[width=78mm]{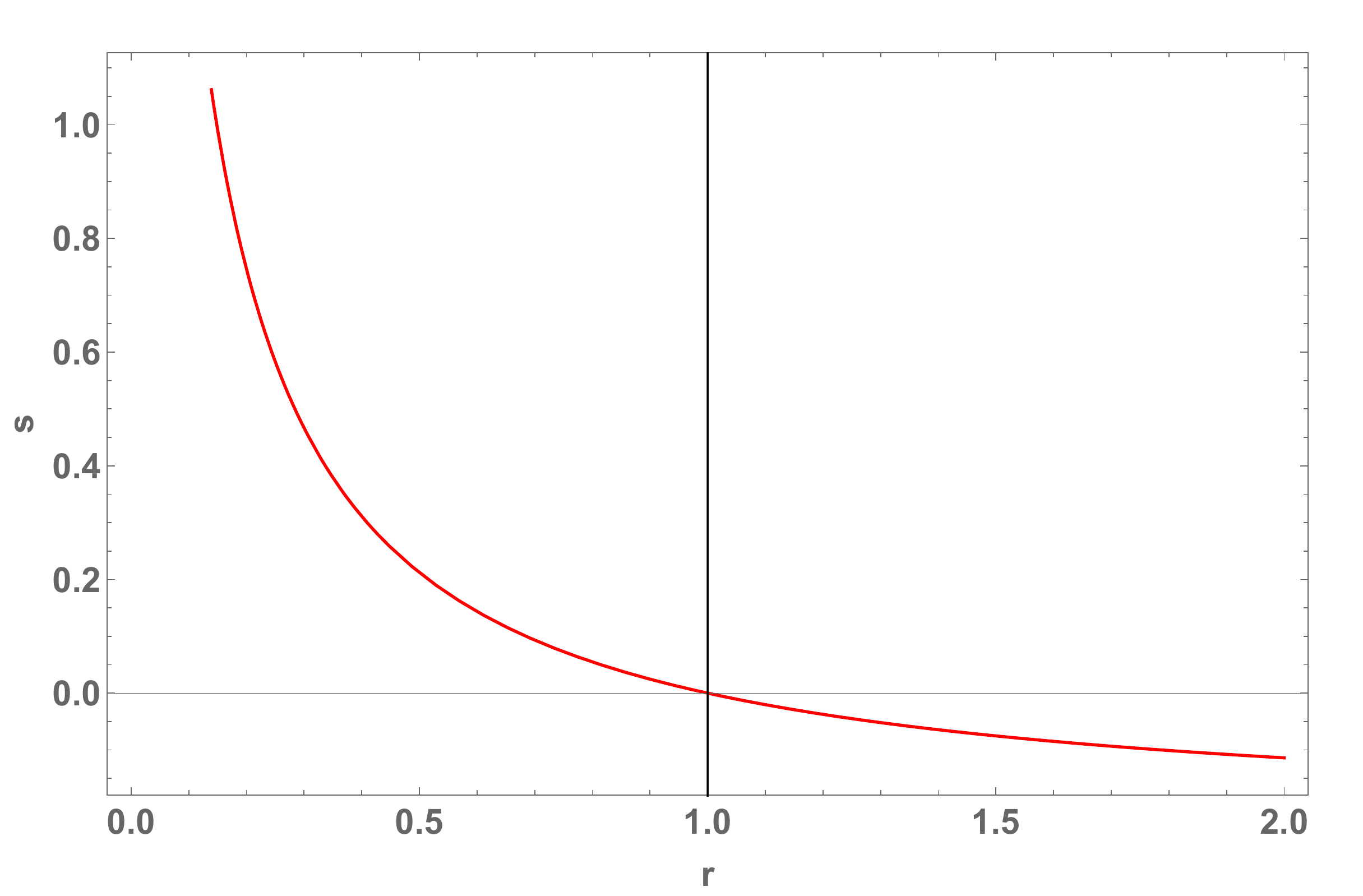}
  \caption{Evolution of r-s parameter with $\beta=0.09804$, $n=1.15$.}\label{fig7}
 \end{figure}

 The matter energy density $(\Omega_m)$ and dark energy density ($\Omega_d$) are obtained as

 \begin{equation}\label{eqn37}
 \Omega_m=\frac{c_1 n^2 \left(\frac{1}{z+1}\right)^{2 n} \left[\left(\left(\frac{1}{z+1}\right)^n\right)^{1/n}\right]^{-3 \omega_m -3}}{3 \beta ^2 \left(\left(\frac{1}{z+1}\right)^{2 n}+1\right)},
 \end{equation} and
 \begin{equation}\label{eqn38}
 \Omega_d=\frac{\left(\left(\frac{1}{z+1}\right)^n\right)^{-\frac{3 (\omega_m +1)}{n}} \left[9 \beta ^2 \left(\left(\frac{1}{z+1}\right)^{2 n}+1\right) \left(\left(\frac{1}{z+1}\right)^n\right)^{\frac{3 (\omega_m +1)}{n}}-4 c_1 n^2 (8 \pi -\alpha  (\omega_m -3)) \left(\frac{1}{z+1}\right)^{2 n}\right]}{24 (\alpha +4 \pi ) \beta ^2 \left(\left(\frac{1}{z+1}\right)^{2 n}+1\right)}.
 \end{equation}
 Adding eqns. (\ref{eqn37}) and (\ref{eqn38}) we get the total energy $(\Omega)$ as
 
\begin{equation}\label{eqn39}
 \Omega= \frac{n^2 \left(\frac{1}{z+1}\right)^{2 n} \left[\frac{c_1 (\alpha  (\omega_m -3)-8 \pi ) \left(\left(\frac{1}{z+1}\right)^n\right)^{-\frac{3 (\omega_m +1)}{n}}+\frac{9 \beta ^2 \left(\left(\frac{1}{z+1}\right)^{-2 n}+1\right)}{4 n^2}}{2 (\alpha +4 \pi )}+c_1 \left(\left(\frac{1}{z+1}\right)^n\right)^{-3/n} \left[\left(\left(\frac{1}{z+1}\right)^n\right)^{1/n}\right]^{-3 \omega_m }\right]}{3 \beta ^2 \left(\left(\frac{1}{z+1}\right)^{2 n}+1\right)}.
\end{equation}
From eqn. (\ref{eqn29}) we find that $q<0$, hence the model represents an accelerating universe. Since
$\displaystyle{\lim_{t \to \infty}}\frac{\sigma^2}{\theta^2}\neq0$ the
universe is anisotropic throughout the evolution. Fig. \ref{fig7} shows the variation of $s$ with respect to
$r$. It is clear from this figure that $s$ is negative when $r$ is
greater than one. As $r\rightarrow\infty$, $s \rightarrow -\infty$
and when $r=1$ we have $s=0$. Hence the universe starts from the
Einstein static era and goes to $\Lambda$CDM era.

\section{Conclusion}
The LRS Bianchi type-I cosmological model in $f(R, T)$ gravity theory is constructed in this paper with the exact solutions of
the field equations. In $f(R, T)$ gravity, cosmic acceleration depends on geometric contribution as well on matter content of the universe. Cosmological models with a source of dark energy yield a very good approximation to the accelerated expansion of the universe. As a result, in this obtained model we can see the behavior of the energy density and pressure with respect to time in Fig. \ref{fig1} and Fig. \ref{fig-3} and with respect to redshift parameter $z$ in Fig. \ref{fig-2} and Fig. \ref{fig-4}. From Fig. \ref{fig1} one can
observe that at initial epoch the energy density of the universe
is very high. As time increases it decreases and approaches to zero
when $t\rightarrow\infty$. Energy density remains positive
throughout the evolution of the universe. From Fig. \ref{fig-3} we see that
the universe starts with a very large negative pressure, decreases
with increase in cosmic time $t$ and approaches to zero for large $t$.
This reveals the characteristic behavior of the dark energy. The nature of
EoS parameter ($\omega_d$) for DE with the evolution of
cosmic time $t$ is shown in Fig. \ref{fig-5} and with respect to redshift parameter $z$ is shown in Fig. \ref{fig-6}. The parameter $\omega>-1$ and $\omega<-1$ correspond quintessence and phantom energy respectively attributes present accelerated expansion of the universe. Also, The evolution of energy density, pressure and EoS parameter correspond to redshift parameter $z$ are depicted in details in the Figs. \ref{fig-2}, \ref{fig-4}, and \ref{fig-6} respectively. 
It can be observed that the universe is dominated by dark energy which may be the strongest evidence for present cosmic expansion. All of the solutions obtained are consistent with the observational results. Hence we feel that these results will be helpful for the researchers to realize the characteristics of the universe in the framework of $f(R, T)$ theory.

\section{Acknowledgements}
The authors would like to thank Prof. P. K. Sahoo for discussion and valuable suggestions. We are grateful to DST, New Delhi, India for providing facilities through DST-FIST lab, Department of Mathematics, where a part of this work was done. The authors also thank the anonymous referee for the comments and suggestions that helped us to improve the manuscript.

\end{document}